\documentclass{ifacconf}
\usepackage{graphicx}      % include this line if your document contains figures
\usepackage{natbib}        	% required for bibliography
\usepackage{amsmath}
\usepackage{amsfonts} 
\usepackage{amssymb}
\usepackage{pdfpages}
\begin{document}
\begin{frontmatter}

\title{Learning stability guarantees for data-driven constrained switching linear systems} 

%\thanks[footnoteinfo]{Sponsor and financial support acknowledgment
%goes here. Paper titles should be written in uppercase and lowercase
%letters, not all uppercase.}

\author[UCL]{Adrien Banse} 
\author[UCL]{Zheming Wang} 
\author[UCL]{Raphaël Jungers}

\address[UCL]{The ICTEAM Institute, UCLouvain, Louvain-la-Neuve,1348,
Belgium (email: adrien.banse@student.uclouvain.be,
zheming.wang@uclouvain.be, raphael.jungers@uclouvain.be).}

\begin{abstract}
We consider stability analysis of constrained switching linear systems in which the dynamics is unknown and whose switching signal is constrained by an automaton. We propose a data-driven Lyapunov framework for providing probabilistic stability guarantees based on data harvested from observations of the system. By generalizing previous results on arbitrary switching linear systems, we show that, by sampling a finite number of observations, we are able to construct an approximate Lyapunov function for the underlying system. Moreover, we show that the entropy of the language accepted by the automaton allows to bound the number of samples needed in order to reach some pre-specified accuracy.
\end{abstract}

\end{frontmatter}
%===============================================================================

\section{Introduction}

In this paper we address the problem of finding probabilistic guarantees for the stability of \emph{constrained switching linear systems} whose dynamics is unknown. 

\textbf{Switching systems.} We consider discrete-time \emph{switching linear systems} (\emph{SLS}) defined by a set $\mathcal{A} = \{ A_i, \}_{i \in \{1, \dots, m\}} $ of $m$ matrices. Their dynamics is given by the following equation: 
\begin{equation}
	x_{t+1} = A_{\sigma(t)} x_t
	\label{dynamics}
\end{equation}
for any $t \in \mathbb{N}$, where $x_t \in \mathbb{R}^n$ and $\sigma(t) \in \{1, \dots, m\}$ are respectively the \emph{state} and the \emph{mode} at time $t$. The sequence $(\sigma(0), \sigma(1), \dots) \subseteq \{1, \dots, m\}^{\mathbb{N}}$ is the \emph{switching sequence}. 

Switching linear systems are an important family of hybrid systems which often arise in Cyber-Physical systems (see \cite{tabuada}). Indeed, the interaction between continuous and discrete dynamics causes hybrid behaviors which makes the stability analysis challenging. In recent years, many model-based stability analysis techniques have been proposed (see \cite{linhai} and references therein, or \cite{jungers_2009_the}).

A \emph{constrained switching linear system} (\emph{CSLS}) is a switching linear system with logical rules on its switching signal. We represent these rules by an \emph{automaton}. The stability of CSLS has also been studied extensively (see e.g. \cite{dai_2011_a}, \cite{PHILIPPE2016242} and \cite{xu2020approximation}). In particular, we are interested in asymptotic stability of CSLS, whose definition is given as follows.
A CSLS whose dynamics is given by \eqref{dynamics} is said to be \emph{asymptotically stable} (or \emph{stable}, for short) if for all $x_0 \in \mathbb{R}^n$,
\begin{equation}
	\lim_{t \to \infty} x_t = 0.
\end{equation}

%Given an automaton $\mathbf{G}$ and a set of matrices \com{$\mathbf{\Sigma} = \{A_i\}_{i \in \{1, \dots, m\}}$, with $m$ the number of matrices}, the system $S(\mathbf{G}, \mathbf{\Sigma})$ is said to be \emph{asymptotically stable} (or \emph{stable}, for short) if\com{, for all switching sequence $(\sigma(0), \sigma(1), \dots)$ accepted by $\mathbf{G}$, and for all initial state $x_0 \in \mathbb{R}^n$,}
%\begin{equation}
%	\com{\lim_{t \to \infty} A_{\sigma(t)} \dots A_{\sigma(0)} x_0 = 0.}
%\end{equation}         
\textbf{Data-driven approach.} In many practical applications, the engineer cannot rely on having a model, but rather has to analyze stability in a \emph{data-driven} fashion. Most classical data-driven methods (see e.g. \cite{2017}, \cite{710876} and \cite{CAMPI200366}) are limited to linear systems and based on classical identification and frequency-domain approaches. These methods may not be well suited for complex systems such as constrained switching linear systems.

In order to tackle hybrid behaviors in switching systems, novel data-driven stability analysis methods have been recently developed based on \emph{scenario optimization} (see \cite{Kenanian2019DataDS}, \cite{berger_2021_chanceconstrained} and \cite{rubbens2021datadriven}). In this paper we seek to take one more step towards complexity. To do that, we develop a data-driven method for providing probabilistic guarantees on the stability of noise-free constrained switching linear systems.

\textbf{Outline.} The rest of this paper is organized in two parts. We introduce the problem that we tackle in Section~\ref{setting}. All concepts needed to this end are introduced in Section~\ref{preli}, and the problem is formulated in Section~\ref{formulation}. In Section~\ref{bind}, we propose a lifting result allowing us to reduce the computation of the \emph{constrained joint spectral radius} to the \emph{joint spectral radius} of a certain set of matrices. Moreover, we state the main theorem of this paper, which extends data-driven results from \cite{berger_2021_chanceconstrained} to constrained switching linear systems. Finally, we investigate further the obtained generalization. We show that the notion of \emph{entropy} can be used to characterize the number of samples needed to reach a specified guarantee on the stability. We will show that, under some assumptions, a smaller entropy allows for a better probabilistic guarantee.

\section{Problem setting}
\label{setting}

\subsection{Preliminaries}
\label{preli}

In this subsection, we introduce the notions necessary to formally present the problem that we solve in this paper.

\textbf{Joint spectral radius.} For arbitrary SLS, given a set of matrices $\mathcal{A} = \{ A_1, \dots, A_m\}$ $\subseteq \mathbb{R}^{n \times n}$, the quantity
\begin{equation}
	\rho(\mathcal{A}) = \lim_{t \to \infty} \max_{\sigma(\cdot) \in \{1, \dots, m\}}  \| A_{\sigma(t-1)} \dots A_{\sigma(0)} \|^{1/t}
	\label{JSRdef}
\end{equation}
is known as the \emph{joint spectral radius} (\emph{JSR}) of a switching linear system defined on $\mathcal{A}$.
The JSR of an switching system rules the stability of the latter:
\begin{prop}[\cite{jungers_2009_the}, Corollary 1.1.] 
Given a set of matrices $\mathcal{A}$, the switching linear system defined by $\mathcal{A}$ is asymptotically stable if and only if $\rho(\mathcal{A}) < 1$.
\end{prop}

It is a well known fact that for any stable arbitrary switching linear system, there is a norm acting as a \emph{common Lyapunov function} (see \cite{jungers_2009_the}, Proposition 1.4.). The following proposition gives a sufficient condition for stability, by restricting the search to \emph{common quadratic Lyapunov functions} (\emph{CQLF}).
\begin{prop}[\cite{jungers_2009_the}, Proposition 2.8] 
\label{approxJSR}
Consider a finite set of matrices $\mathcal{A}$. If there exists $\gamma \geq 0$ and a symmetric matrix $P \succ 0$ such that $A^TPA \preceq \gamma^2 P$ holds for any matrix $A \in \mathcal{A}$, then $\rho(\mathcal{A}) \leq \gamma$.
\end{prop}

\textbf{Constrained joint spectral radius.} First, we give the definition of an \emph{automaton}. An automaton is a strongly connected, directed and labelled graph $\mathbf{G}(V, E)$ with $V$ the set of nodes and $E$ the set of edges. Note that we drop the explicit writing of $V$ and $E$ when it is clear from the context. The edge $(v, w, \sigma) \in E$ between two nodes $v, w \in V$ carries the \emph{label} $\sigma \in \{1, \dots, m\}$, which maps to a mode of the switching system. A sequence of labels $(\sigma(0), \sigma(1), \dots)$ is a \emph{word} in the language \emph{accepted} by the automaton $\mathbf{G}$ if there is a path in $\mathbf{G}$ carrying the sequence as the succession of the labels on its edges. A CSLS defined on the set of matrices $\mathbf{\Sigma}$ and constrained by the automaton $\mathbf{G}$ is noted $S(\mathbf{G}, \mathbf{\Sigma})$. We define the set of all possible products of matrices in $\mathbf{\Sigma}$ of length $l$ given an automaton $\mathbf{G}$ as
\begin{equation}
\begin{aligned}
    &\mathbf{\Pi}_l = \{ A_{\sigma(l-1)} A_{\sigma(l-2)} \dots A_{\sigma(0)} :   \\
    & \quad  (\sigma(0), \sigma(1), \dots, \sigma(l-1)) \text{ is a word of }\mathbf{G} \}.
\end{aligned} 
\end{equation}
The constrained joint spectral radius (\emph{CJSR}), which is a generalization of the JSR to CSLS, was first introduced in \cite{dai_2011_a}. Given a set of matrices $\mathbf{\Sigma}$ and an automaton $\mathbf{G}$, the CJSR of the constrained switching linear system $S(\mathbf{G}, \mathbf{\Sigma})$ is defined as 
\begin{equation}
	\rho(\mathbf{G}, \mathbf{\Sigma}) = \lim_{t \to \infty} \max \left\{ \| \mathbf{A} \|^{1/t} \, : \, \mathbf{A} \in \mathbf{\Pi}_t \right\}.
	\label{CJSRdef}
\end{equation}
In the same way, the stability of a constrained switching linear system is characterized by its CJSR:
\begin{prop}[\cite{dai_2011_a}, Corollary 2.8.]
\label{daiStability}
Given a set of matrices $\mathbf{\Sigma}$ and an automaton $\mathbf{G}$, the constrained switching linear system $S(\mathbf{G}, \mathbf{\Sigma})$ is asymptotically stable if and only if $\rho(\mathbf{G}, \mathbf{\Sigma}) < 1$.
\end{prop}

\subsection{Problem formulation}
\label{formulation}

We will now formally present the problem that we solve in this paper. 

\textbf{Model-based setting.} Consider a given constrained switching linear system $S(\mathbf{G}, \mathbf{\Sigma})$ with $\mathbf{\Sigma} \subseteq \mathbb{R}^{n \times n}$. Let $\Delta = \mathbb{S} \times \mathbf{\Pi}_l$  with $S \subseteq \mathbb{R}^n$ the unit sphere and $\mathbf{\Pi}_l$ the set of all admissible products of length $l$. We introduce the following optimization problem\footnote{We note $\min(f(x),g(x))$ the multiobjective optimization problem where g(x) is used as a \emph{tie-breaking rule}. That is, the objective is to minimize the function $f(x)$, and, in case there are several optimizers, the solution is the one which minimizes $g(x)$. Observe that the latter is unique because the problem is quasi-convex, and because $\| \cdot \|$ is a strongly convex function.} (see \cite{berger_2021_chanceconstrained}):
\begin{equation}
\begin{split}
    \mathcal{P}(\Delta):  &\min_{\substack{P \in \mathbb{R}^{n \times n} \\ \gamma \geq 0}} (\gamma, ||P||^2_F) \\ 
            		 \text{ s.t. }   &P \in \mathcal{X} := \left\{ P : I \preceq P \preceq CI  , \, P = P^T \right\},  \\
                          &(\mathbf{A}x)^TP(\mathbf{A}x) \leq \gamma^{2l} x^T P x \quad \forall (x, \mathbf{A}) \in \Delta,
\end{split}
\label{cqlfINF}
\end{equation}
for a large $C \in \mathbb{R}_{\geq 0}$, where $\| \cdot \|_F$ is the Frobenius norm. We denote $(\gamma^*(\Delta), P^*(\Delta))$ as the solution of optimization problem \eqref{cqlfINF}.

Following Proposition~\ref{approxJSR}, Program~\eqref{cqlfINF} allows us to study stability in a model-based setting i.e., when $\Delta$ is known. Indeed if $\gamma < 1$, then the ellipsoidal norm $\| \cdot \|_{P^*(\Delta)}$ is a CQLF for the considered CSLS \cite[]{jungers_2009_the}. Observe that, in addition to the problem of Proposition~\ref{approxJSR}, a tie-breaking rule is defined in Program~\eqref{cqlfINF}. This tie-breaking rule allows for improving the probabilistic guarantees we obtain in Theorem~\ref{main} (see \cite{Kenanian2019DataDS} for details). A constraint $P \preceq CI$ is also added to ensure that the set of feasible $P$ is compact, so that the existence of a solution is guaranteed\footnote{For more details about these additions, see \cite{berger_2021_chanceconstrained}.}.

\textbf{Data-driven setting.} In this work, we analyze the same problem in a data-driven framework: we assume that the system is not known (i.e., $\mathbf{A}$ is not known in Program~\eqref{cqlfINF}), but that we sample $N$ trajectories of length $l$ of a system $S(\mathbf{G}, \mathbf{\Sigma})$. The $i$-th trajectory is noted $(x_{i, 0}, \dots, x_{i, l})$ for $i \in \{1, \dots, N\}$. The trajectories are assumed to be generated from initial states $x_{i, 0}$ drawn randomly, uniformly and independently from $\mathbb{S}$, the unit sphere.

For each trajectory $i \in \{1, \dots, N\}$, the $l$ matrices are generated from the automaton $\mathbf{G}(V, E)$ in the following way. An initial state $u_0$ is drawn randomly and uniformly from $V$. Then a random walk of length $l$ is performed on $\mathbf{G}$, where, from $u_j \in V$, the next state $u_{j+1}$ is drawn randomly, uniformly and independently from the set of its out-neighbours $\left\{ u_{j+1} \in V : (u_j, u_{j+1}, \sigma_i(j) \in E \right\}$ where $\sigma_i(j)$ is the label corresponding to the edge linking $u_j$ and $u_{j+1}$. The sequence of nodes $(u_0, \dots, u_j,u_{j+1}, \dots, u_l)$ form a switching sequence $\sigma_i(0), \dots, \sigma_i(l - 1)$, which maps to the matrices $A_{\sigma_i(0)}, \dots, A_{\sigma_i(l-1)}$.

We define the set of $N$ observations $\omega_N$ as
\begin{equation}
    \omega_N = \{ (x_{i, 0}, \mathbf{A}_i), i = 1, \dots, N \}
\end{equation}
where $\mathbf{A}_i = A_{\sigma_i(l-1)} \dots A_{\sigma_i(0)} \in \mathbf{\Pi}_l$. Note that the observations in $\omega_N$ are assumed to be noise-free.

We define $\mathbb{P} = \mathbb{P}_x \times \mathbb{P}_\sigma$ the probability measure on $\Delta$ with $\mathbb{P}_x$ the uniform distribution on $\mathbb{S}$ and $\mathbb{P}_\sigma$ the probability distribution describing the distribution of paths in $\mathbf{\Pi}_l$ as explained above. Note that $\mathbb{P}_\sigma$ is not necessarily a uniform measure.

Now, for a given set $\omega_N$, let us define the \emph{sampled optimization problem} $\mathcal{P}(\omega_N)$ associated to $\mathcal{P}$:
\begin{equation}
\begin{split}
    \mathcal{P}(\omega_N) :  &\min_{\substack{P \in \mathbb{R}^{n \times n} \\ \gamma \geq 0}} (\gamma, ||P||^2_F)  \\
            \text{ s.t. }  &P \in \mathcal{X} := \left\{ P : I \preceq P \preceq CI , \, P = P^T \right\},  \\
                            & (\mathbf{A}x)^TP(\mathbf{A}x) \leq \gamma^{2l} x^T P x \quad \forall (x, \mathbf{A}) \in \omega_N,
\end{split}
\label{cqlfSAMPLED}
\end{equation}
We denote $(\gamma^*(\omega_N), P^*(\omega_N))$ as the solution of optimization problem~\eqref{cqlfSAMPLED}, and $\text{Cost}(\omega_N)$ its optimal cost. The problem $\mathcal{P}(\omega_N)$ defined in Program~\eqref{cqlfINF} is the \emph{data-driven} version of the optimization problem $\mathcal{P}(\Delta)$  defined in Program~\eqref{cqlfSAMPLED}. The issue that we tackle in this paper is the inference of $\gamma^*(\Delta)$, the solution of optimization problem~\eqref{cqlfINF} from $(\gamma^*(\omega_N), P^*(\omega_N))$ with a certain user-defined level of confidence.

\section{Main results}
\label{bind}

In this section, we present our main results. First, in Proposition~\ref{reduction} given an automaton $\mathbf{G}$ and a set of matrices $\mathbf{\Sigma}$, we show that the CJSR can be bounded by the classical JSR of the set of all admissible products of a given length $\mathbf{\Pi}_l$. Even though other reductions of the CJSR computation problems to a simpler JSR have already been proposed in the literature (see e.g. \cite{dai_2011_a} and \cite{PHILIPPE2016242}), to the best of our knowledge, Proposition~\ref{reduction} is new, and will be useful for our purposes. Second, we use this result in order to derive a probabilistic guarantee allowing to relate the data-driven problem~\eqref{cqlfSAMPLED} to the model-based problem~\eqref{cqlfINF}. This guarantee is given in Theorem~\ref{main}.

\begin{prop}
\label{reduction}
For all $l > 0$, given an automaton $\mathbf{G}$ and a set of matrices $\mathbf{\Sigma}$, the CJSR of $S(\mathbf{G}, \mathbf{\Sigma})$ and the JSR of the switching linear system defined by $\mathbf{\Pi}_l$ satisfy
\begin{equation}
	\rho(\mathbf{G}, \mathbf{\Sigma}) \leq \rho(\mathbf{\Pi}_l)^{1/l}.
	\label{finiteIneq}
\end{equation}
Moreover, the equality holds asymptotically i.e.,
\begin{equation}
\label{asymEq}
	\rho(\mathbf{G}, \mathbf{\Sigma}) = \lim_{l \to \infty} \rho(\mathbf{\Pi}_l)^{1/l}.
\end{equation}
\end{prop}
Proposition~\ref{reduction} allows us to reduce the problem of approximating the CJSR to the problem of approximating the JSR of another arbitrary switching linear system. Therefore we can generalize previous data-driven works on arbitrary systems. In particular, we draw our results on top of \cite{berger_2021_chanceconstrained} in order to obtain data-driven stability guarantees for constrained systems.

We remark that the data-driven problem~\eqref{cqlfSAMPLED} is a quasi-linear optimization problem, as defined in \cite[Equation 1]{berger_2021_chanceconstrained}. Thus, a very similar analysis as in \cite{berger_2021_chanceconstrained}, based on scenario-approach results \cite{calafiore_2010}, can be done. First, we recall the definition of a \emph{Barabanov} matrix (see \cite{berger_2021_chanceconstrained}, Defintion 7). A matrix $A \in \mathbb{R}^{n \times n}$ is said to be Barabanov if there exists a symmetric matrix $P \succ 0$ and $\gamma \geq 0$ such that $A^TPA = \gamma^2P$.

Given Proposition~\ref{daiStability}, the following theorem generalizes Corollary 14 of \cite{berger_2021_chanceconstrained}. It gives probabilistic guarantees for the stability of a constrained switching linear system. In the following theorem, $\Phi( \cdot, a, b)$ denotes the \emph{regularized incomplete beta function} for the two parameters $a, b \in \mathbb{N}$  (see \cite{Kenanian2019DataDS}, Definition 2).

\begin{thm}
\label{main}
Consider an automaton $\mathbf{G}$, a set of matrices $\mathbf{\Sigma} \subseteq \mathbb{R}^{n \times n}$, samples $\omega_N \subset \Delta$ obtained as explained in Section~\ref{formulation}, a fixed length $l > 0$ and $N \geq d := n(n+1)/2$. Suppose $\mathbf{\Pi}_l$ contains no Barabanov matrices. Consider problem $P(\omega_N)$ with solutions $\gamma^*(\omega_N)$ and $P^*(\omega_N)$. Then, for a given level of confidence $\beta \in (0, 1)$, 
\begin{equation}
	\mathbb{P}\left( \left\{ \omega_N \in \Delta^N : \, \rho(\mathbf{G}, \mathbf{\Sigma})
	\leq \frac{\gamma^*(\omega_N)}{\sqrt[l]{\delta(\beta, \omega_N)}} \right\} \right)
	\geq \beta,
\label{upperBoundMTNS} 
\end{equation}
and the function $\delta(\beta, \omega_N)$ takes the form
\begin{equation}
\sqrt{1 - \Phi^{-1}(\varepsilon(\beta, N)\kappa(P^*(\omega_N)) / p_{l, \min} , (n-1)/2, 1/2)}.
\end{equation}
where $p_{l, \min}$ is the minimal probability of all matrices in $\mathbf{\Pi}_l$, $\kappa(P) = \sqrt{\det(P)/\lambda_{\min}(P)^n}$, and $\varepsilon(\beta, N)$ takes the closed form 
\begin{equation}
	\varepsilon(\beta, N) = 1 - \Phi(1-\beta, d+1, N-d).
\end{equation}
\end{thm}

Theorem~\ref{main} provides a general way of obtaining probabilistic stability guarantees. Indeed, for a given confidence level $\beta$, if one computes an upper bound~\eqref{upperBoundMTNS} strictly less than 1, then following Proposition~\ref{daiStability}, stability holds with probability at least $\beta$. 

We now show how one can use it in practice, by deriving a few corollaries. The following corollary holds if the distribution of drawing a product in $\mathbf{\Pi}_l$ is uniform.

\begin{cor}
\label{uniform}
Suppose $\mathbb{P}_\sigma$ is a uniform measure. Then the function $\delta(\beta, \omega_N)$ in Theorem~\ref{main} can be written
\begin{equation}
\sqrt{1 - \Phi^{-1}(\varepsilon(\beta, N) | \mathbf{\Pi}_l | \kappa(P^*(\omega_N)), (n-1)/2, 1/2)}.
\end{equation}
\end{cor}

We now show that we can push further our analysis of the upper bound expressed in Corollary~\ref{uniform} by using the notion of \emph{entropy} \cite[Definition 4.1.1]{lind_marcus_1995}. Let $| \mathcal{L}_{\mathbf{G}, l} |$ be the language accepted by $\mathbf{G}$ restricted to length $l$. The entropy $h(\mathbf{G})$ of $\mathbf{G}$ is the growth rate of $| \mathcal{L}_{\mathbf{G}, l} |$ i.e.,
\begin{equation}
	h(\mathbf{G}) = \lim_{l \to \infty} \frac{\log_2|\mathcal{L}_{\mathbf{G}, l}|}{l}.
\end{equation}

\begin{figure}
	\includegraphics[width = \linewidth]{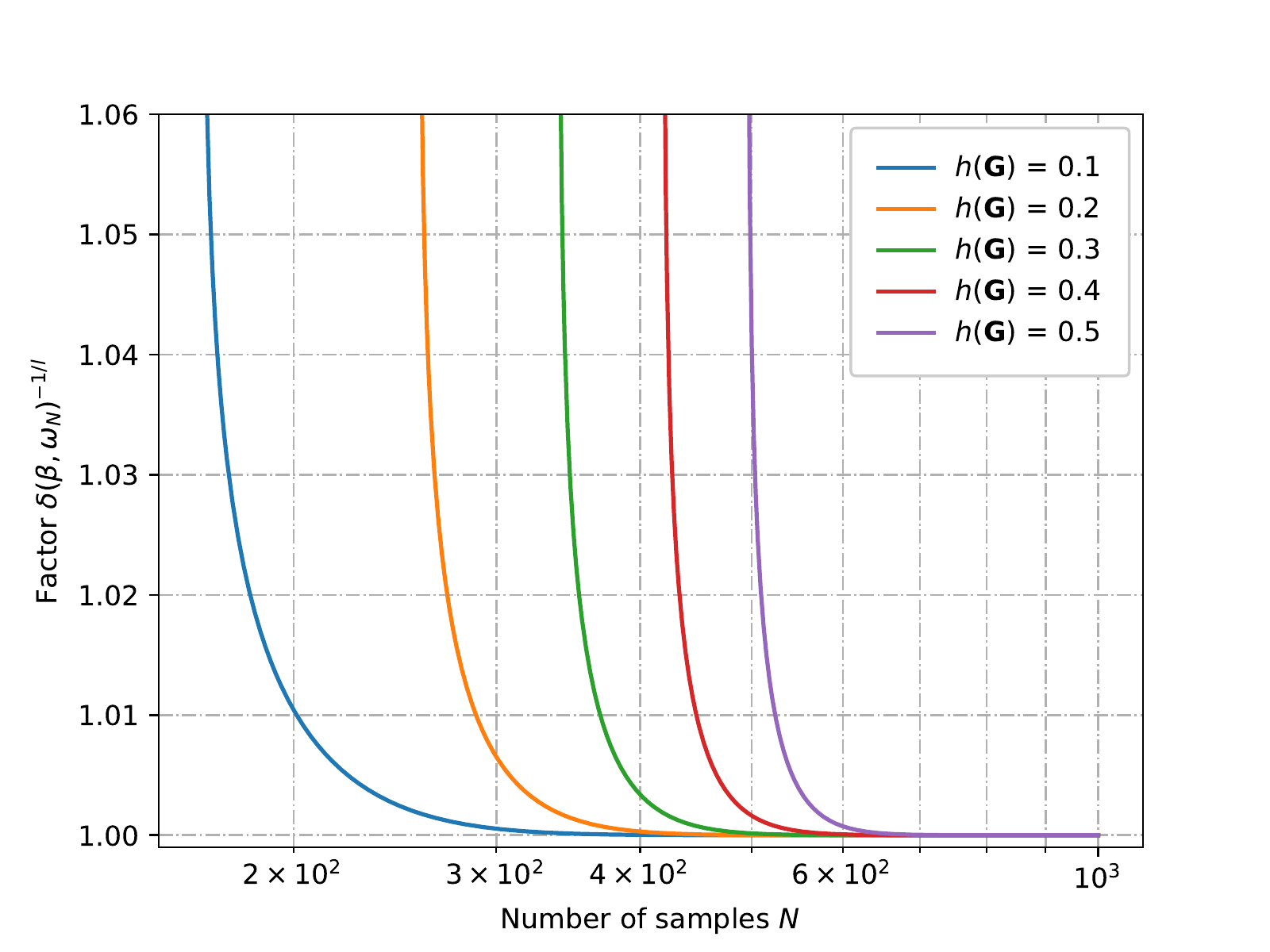}
	\caption{Shape of the factor $1/\sqrt[l]{\delta(\beta, \omega_N)}$ in Theorem~\ref{main} with respect to the entropy, for a confidence level $\beta = 95\%$, a large $l$ (here $l = 50$) and $n = 2$. One can see that this factor converges to 1 as $N$ increases, and that a smaller entropy allows to converge faster.}
	\label{MTNSplot}
\end{figure}

Since $|\mathbf{\Pi}_l | \leq | \mathcal{L}_{\mathbf{G}, l} |$, the definition of the entropy gives the following corollary. 
\begin{cor}
\label{asymDelta}
For $l \to \infty$, the function $\delta(\beta, \omega_N)$ in Corollary~\ref{uniform} can be written
\begin{equation}
\lim_{l \to \infty} \sqrt{1 - \Phi^{-1}(\varepsilon(\beta, N) 2^{lh(\mathbf{G})} \kappa(P^*(\omega_N)), (n-1)/2, 1/2)}.
\end{equation}
\end{cor}
Corollary~\ref{asymDelta} provides an asymptotic estimate of the probabilistic upper bound in Theorem~\ref{main}, as a function of the entropy of the automaton $\mathbf{G}$. One can see that an automaton with small entropy allows for a better estimate of the CJSR, for a fixed number of samples. This is illustrated in Figure~\ref{MTNSplot}.

Now we show that we can also derive a practical bound for any finite $l>0$, unlike Corollary~\ref{asymDelta} which holds asymptotically. For this we use classical results from graph theory.

\begin{prop}
\label{graphTheory}
Let $A$ be the adjacency matrix of some automaton $\mathbf{G}(V, E)$. Let $\lambda_1 \leq \dots \leq \lambda_{|V|}$ be the eigenvalues of $A$. Assume $A$ is diagonalizable. Then for any $l \geq 0$, $|\mathbf{\Pi}_l| \leq |V| \lambda_n^l$.
\end{prop} 

Proposition~\ref{graphTheory} directly gives the following corollary.
\begin{cor}
\label{finiteDelta}
	Let $A$ be the adjacency matrix of some automaton $\mathbf{G}(V, E)$ Let $\lambda_1 \leq \dots \leq \lambda_{|V|}$ be the eigenvalues of $A$. Assume $A$ is diagonalizable. Then for any $l > 0$, the function $\delta(\beta, \omega_N)$ in Corollary~\ref{uniform} can be written 
\begin{equation}
	\sqrt{
	1 - \Phi^{-1}(\varepsilon(\beta, N) |V| \lambda_n^{l} \kappa(P^*(\omega_N)), (n-1)/2, 1/2)
	}.
\end{equation}
\end{cor}

Corollary~\ref{finiteDelta} provides a probabilistic upper bound in Theorem~\ref{main}, as a function of the largest eigenvalue of the adjacency matrix of $\mathbf{G}$. One can see that an automaton with a small largest eigenvalue allows for a better estimate of the CJSR, for a fixed number of samples $N$ and length $l$.

\section{Conclusion}

In this work, we extended the scope of data-driven stability analysis of hybrid systems by generalizing previous data-driven results to the constrained case. In particular we have built our results on the basis of \cite{berger_2021_chanceconstrained}.

We proceeded as follows. We first proposed a lifting result allowing us to reduce the computation of the CJSR of a given CSLS to the computation of a simpler JSR. We then stated the main theorem of this paper, which provides probabilistic guarantees for the stability of a given noise-free CSLS. Finally, we claimed that in case of uniformity on the distribution of switching sequences, we can investigate further the obtained bound. We showed that a smaller entropy of the automaton allows for a better guarantee about the stability.

In further research, we plan to extend this type of method to noisy observations. We also plan to investigate different approaches. For example, getting rid of the lifting result would allow to reduce the conservativism introduced by the latter, i.e. the gap between the lifted JSR $\rho(\mathbf{\Pi}_l)^{1/l}$ and the true CJSR $\rho(\mathbf{G}, \mathbf{\Sigma})$ in \eqref{finiteIneq}. In this regard we plan to directly approximate \emph{multiple Lyapunov functions} \cite[Definition~2]{multinorm}.

\bibliography{ifacconf}             
                                                  
\end{document}